# Anomalous high-pressure Jahn-Teller behavior in CuWO$_4$

J. Ruiz-Fuertes[1,2,*], A. Segura[1,2], F. Rodríguez[1,3], D. Errandonea[1,2], and M. N. Sanz-Ortiz[1,3]

[1]MALTA Consolider Team

[2]Departamento de Física Aplicada – ICMUV, Universitat de València, Edificio de Investigación, c/ Dr. Moliner 50, 46100 Burjassot, Valencia, Spain

[3]DCITIMAC, Facultad de Ciencias, Universidad de Cantabria, 39005 Santander, Spain

**Abstract:** High-pressure optical-absorption measurements performed in CuWO$_4$ up to 20 GPa provide experimental evidence of the persistence of the Jahn-Teller (JT) distortion in the whole pressure range both in the low-pressure triclinic and in the high-pressure monoclinic phase. The electron-lattice coupling associated with the $e_g$(E⊗e) and $t_{2g}$(T⊗e) orbitals of Cu$^{2+}$ in CuWO$_4$ are obtained from correlations between the JT distortion of the CuO$_6$ octahedron and the associated structure of Cu$^{2+}$ $d$-electronic levels. This distortion and its associated JT energy ($E_{JT}$) decrease upon compression in both phases. However, both the distortion and associated $E_{JT}$ increase sharply at the phase transition pressure ($P_{PT}$ = 9.9 GPa) and we estimate that the JT distortion persists for a wide pressure range not being suppressed up to 37 GPa. These results shed light on the transition mechanism of multiferroic CuWO$_4$ suggesting that the pressure-induced structural phase transition is a way to minimize the distortive effects associated with the toughness of the JT distortion.

[*] Corresponding author: Email: javier.ruiz-fuertes@uv.es, Tel.: (34) 96 354 4908, Fax: (34) 96 354 3146

Electron-lattice coupling is one of the fundamental issues to understand a wide variety of relevant physical phenomena in materials science [1-6]. In particular, the E⊗e Jahn-Teller (JT) effect, involving orbitally degenerate $e_g$ electrons and lattice distortions (or vibrations) of $e_g$ symmetry, is known to play a crucial role in many physical phenomena of transition-metal oxides such as colossal magnetoresistance [7], insulator to metal transition [8, 9], or spin transition phenomena [10, 11]. The study of electron-lattice coupling under high pressure conditions has become a key topic in condensed matter physics [4] since the discovery of the increase of critical temperature in $Ba_2LaCu_3O_{7-y}$ under compression [3]. In general, the knowledge of how electron-lattice couplings and their associated phenomena behave in compound series and under compression has received a lot of attention in the last decade.

In octahedral $Cu^{2+}$ or $Mn^{3+}$ oxides, JT effect yields low-symmetry distortion around the transition-metal ion increasing the stabilization energy ($E_{JT}$) [5, 10]. Basic models predict the reduction of the JT distortion upon compression due to the hardening of the coupled vibration or the relative weakening of the electron-lattice coupling induced by electron delocalization. Electron-lattice coupling can be eventually suppressed under high-pressure conditions leading to the JT distortion quenching [8-10]. The lack of distortion in the metallic phase of many transition-metal oxides is usually associated with such suppression along with the insulator-to-metal transitions ($LaMnO_3$). The pressure dependences of $E_{JT}$ and JT distortion, $\rho = \sqrt{Q_\theta^2 + Q_\varepsilon^2}$, described in terms of the tetragonal and rhombic normal coordinates ($Q_\theta$, $Q_\varepsilon$) for strongly coupled $d^4$ and $d^9$ transition-metal systems can be found elsewhere [10-12]. According to estimates based on the volume dependence of electron-lattice coupling and vibrational energy of the coupled mode, it is unlikely that a static E⊗e JT distortion increases after volume compression along a pressure-induced phase transition.

However, recent high-pressure structural studies in $CuWO_4$ found the opposite behavior [13]. We will show how such an unusual behavior is a consequence of the reaction of the $CuO_6$ octahedron against the JT strength. Electronic and crystal structure correlations performed in this work suggest that the triclinic-to-monoclinic phase transition in $CuWO_4$ is mainly driven by reorientations of the $CuO_6$ octahedra towards easier-distortion directions as a way to preserve the JT distortion. Thus this system constitutes a model example of cooperative JT-driven structural phase transition induced by pressure.

Here, we investigate the *d*-electron structure associated with $Cu^{2+}$ in $CuWO_4$ by high-pressure optical absorption spectroscopy to establish structural correlations yielding first experimental electron-lattice coupling determination in $Cu^{2+}$. We aim to find the relation between $E_{JT}$ and $\rho \approx Q_\theta$ ($Q_\varepsilon \approx 0$), the pressure dependence of which is known from x-ray diffraction (XRD) [13].

$CuWO_4$ crystallizes in a triclinic ($P\bar{1}$) phase at ambient conditions (Fig. 1a), and is antiferromagnetic below $T_N$ = 23 K. It undergoes a structural phase transition at 10 GPa, to a monoclinic (*P2/c*) wolframite-type structure. The phase transition involves an abrupt reorientation of the $CuO_6$ octahedra that remain highly distorted in the high pressure phase.

The variation of the optical-absorption spectrum of $CuWO_4$ with pressure, both for the triclinic and monoclinic phases, is shown in Fig. 2. We used (010) cleavage single crystals with thicknesses from 10 to 20 μm that were loaded together with a ruby chip into a 40 μm-thickness, 250 μm-diameter hydrostatic cavity placed between two 500 μm-culet diamonds of a membrane-type anvil cell. Both methanol-ethanol-water (16:3:1) and silicone oil were used as pressure-transmitting media. The spectroscopy setup is described elsewhere [14, 15].

The absorption spectra of CuWO$_4$ (Fig. 2) can be explained on the basis of Cu$^{2+}$ $d$-$d$ intraconfigurational transitions within a JT-distorted CuO$_6$ with pseudo-elongated $D_{4h}$ coordination (Fig. 1b). The absorption bands basically correspond to electronic transitions from the parent octahedral $t_{2g}$ and $e_g$ filled orbitals: $e_g$, $b_{2g}$ and $a_{1g}$, to the singly occupied $b_{1g}$ orbital (Fig. 1c). These transitions, which are usually weak in centrosymmetric systems, appear enhanced in CuWO$_4$ by both non-centrosymmetric crystal-field distortions and the exchange mechanism [16]. The two broad bands observed in the low-pressure phase ($P$ < 9.9 GPa) correspond to $E_1$, $E_2$, and $E_3$ following the scheme of Fig. 1c. As usually observed in Cu$^{2+}$ oxides, the two crystal-field transitions associated with the $t_{2g}$-octahedral orbitals, overlap making their assignment difficult [17]. However, these bands are resolved by polarized absorption spectroscopy (Fig. 3). Due to the distinct band intensity shown by the absorption spectrum in each polarization, we have derived a difference spectrum, which only contains the two higher energy transitions. Hence, we assign $E_1$ = 1.16 eV to $a_{1g} \rightarrow b_{1g}$, $E_2$ = 1.34 eV to $b_{2g} \rightarrow b_{1g}$ and $E_3$ = 1.56 eV to $e_g \rightarrow b_{1g}$ at ambient pressure. The corresponding tetragonal splitting of $t_{2g}$ and $e_g$ octahedral orbitals are $\Delta_e$ = $E_1$ = 1.16 eV and, $\Delta_t$ = $E_3$ - $E_2$ = 0.22 eV. Upon compression, $E_1$, $E_2$ and $E_3$ shift to lower energies up to 9.4 GPa. Accordingly, $\Delta_e$ and $\Delta_t$ decrease with pressure, and correlate with the reduction of the JT distortion observed by XAS and XRD [13] (Fig. 4). At $P$ > 9.9 GPa, the spectrum abruptly changes and an additional narrow band appears at 1.19 eV, this change in the electronic structure is due to the pressure-induced triclinic-to-monoclinic phase transition [13, 18, 19]. According to XRD and XAS [13], the Cu$^{2+}$ elongated low-symmetry coordination remains in the high-pressure phase, as it is confirmed by the distinctive $d$-splitting pattern related to the JT distortion. The magnitude of the pseudo-tetragonal distortion sharply increases and the O-Cu-O elongation changes from one direction to another at the phase transition (Fig. 1a). This structural change is also detected by Raman spectroscopy through the abrupt redshift in the Cu-O-related mode at 316 cm$^{-1}$ ($P = 0$ GPa) and the different pressure coefficient found for each phase [19].

Figure 4 shows the variation of $E_1$, $E_2$ and $E_3$ with pressure for the two phases and the corresponding variations of $\Delta_e$ and $\Delta_t$ as a function of the JT distortion $\rho = Q_\theta$ ($Q_\varepsilon \approx 0$) of the CuO$_6$ octahedron. In the low-pressure phase, $E_3$ redshifts at a rate of -11 meV GPa$^{-1}$ and the masked $E_2$ shifts -12 meV GPa$^{-1}$. However $E_1$ exhibits a pronounced redshift of -34 meV GPa$^{-1}$ which correlates with the decrease with pressure of the CuO$_6$ distortion derived from XRD. The sudden increase experienced by all $E_1$, $E_2$ and, $E_3$ and $Q_\theta$ ($Q_\varepsilon \approx 0$) at 9.9 GPa shows the phase-transition onset. In the high-pressure monoclinic phase, $E_1$ slightly decreases with pressure but $E_2$ and $E_3$ increase in such a way that $\Delta_e$ and $\Delta_t$ decrease with pressure (increase with $Q_\theta$). These findings are noteworthy as they provide the dependences of $\Delta_e$ and $\Delta_t$ with $P$ and $Q_\theta$ which are directly related to the electron-lattice coupling through $\partial \Delta_{e,t}/\partial Q_\theta$. In the low-pressure triclinic phase the $e_g$ splitting depends linearly with $Q_\theta$ as $\Delta_e = K_e Q_\theta$, with $K_e = 2.3$ eV Å$^{-1}$. The same dependence is found for $\Delta_t = K_t Q_\theta$, although a different coupling coefficient is measured in each phase: $K_t = 0.5$ and 1.2 eV Å$^{-1}$ for the triclinic and monoclinic phases, respectively. These values are in good agreement with previous structural correlations in Cu$^{2+}$ compound series involving CuCl$_6$ and CuF$_6$ [11, 17], with $K_e$ and $K_t$ values of 2.4 and 0.4 eV Å$^{-1}$, respectively. It must be noted that the electron-lattice coupling coefficient in the JT theory, named $A_e$ in E⊗e ($A_t$ in T⊗e), is related to the JT splitting derivative as $A_e = 1/2\ \partial \Delta_e/\partial Q_\theta$ ($A_t = 2/3\ \partial \Delta_t/\partial Q_\theta$) [20]. In E⊗e (or T⊗e) model, $A_e$ (or $A_t$) and the force constant of the corresponding lattice mode determine stabilization energy and the corresponding JT distortion of the CuO$_6$ octahedron [18, 20].

Following EXAFS and XRD results [13] we deduce that the axial bond length variation in the low-pressure phase $\partial R_{ax}/\partial P$ = -0.016 Å GPa$^{-1}$, is an order of magnitude larger than the equatorial one $\partial R_{eq}/\partial P$ = -0.002 Å GPa$^{-1}$. According to this, pressure-induced suppression of the JT distortion should occur at about 37 GPa. A similar estimate is obtained by extrapolating structural data from the high-pressure monoclinic phase, thus suggesting that pressure-induced JT quenching towards a regular octahedron is mainly governed by the JT stabilization energy of the CuO$_6$ rather than the particular crystal phase to which it belongs. In fact, a similar quenching pressure was derived for CuCl$_6$ in Rb$_2$CuCl$_4$ [12], where the JT energy, $E_{JT} = E_1/4 = 0.3$ eV, is similar to the measured in CuWO$_4$. This feature envisages the difficulty to suppress the JT distortion in Cu$^{2+}$ systems, for which severe pressure conditions are required to overpass $E_{JT}$. The high stability of the CuO$_6$ distortion, persisting below 37 GPa in CuWO$_4$, forces the axial O-Cu-O to flip direction towards easier-distortion paths as a way to react against compression. This re-accommodation of the CuO$_6$ octahedron along easy-distortion directions may be considered the origin of the phase transition, we thus suggest that the JT effect is the main driving force triggering the structural transition. This proposed scenario correlates with spectroscopic results. The rapid $E_{JT}$ decrease with pressure in the triclinic phase (-90 meV in 10 GPa) involves an important anisotropy stress along the CuO$_6$ axial distortion, which is eventually released in the triclinic-to-monoclinic phase transition, as it is reflected by the unexpected increase of $Q_\theta$ and $E_{JT}$ observed at the phase-transition pressure (Fig. 4).

In conclusion, optical absorption of CuWO$_4$ unravels the JT distortion and associated $E_{JT}$ through the characteristic $d$-orbital splitting pattern of Cu$^{2+}$ measured as a function of pressure up to 20 GPa. Pressure continuously reduces the JT distortion and $E_{JT}$ of CuO$_6$ up to the triclinic-to-monoclinic phase transition at 9.9 GPa. Contrary to expectations, both the distortion and JT energy abruptly increase at the phase transition, what is a quite unconventional phenomenon. The reaction of $CuO_6$ octahedron against the reduction of the JT distortion suggests that the JT effect is the main driving force triggering the phase transition. The pressure-induced structural variations modify the exchange paths between $Cu^{2+}$ ions yielding a change in the $CuWO_4$ magnetic behavior from antiferromagnetic to ferromagnetic as it has been recently suggested from *ab initio* calculations [19]. Accordingly, present results constitute experimental support for understanding the changes of magnetic properties through structural transformations mediated by the JT effect.

**Acknowledgements**

C.-Y. Tu is acknowledged for providing us the crystals used to perform the experiments. J.R.F. is indebted to the FPI research grant (BES-2008-002043) and thanks C. Renero-Lecuna for fruitful discussions on the spectroscopic data. The authors thank the financial support from the MICINN of Spain under grants No. MAT2010-21270-C04-01, MAT2008-06873-C02-01/02 and CSD2007-00045.

**Figures captions:**

**Figure 1.** (color online). (a) Crystal structure of CuWO$_4$ low-pressure triclinic (left) and high-pressure monoclinic (right) phases. (b) CuO$_6$ octahedral simplified sketch showing the pseudo-elongated $D_{4h}$ symmetry. The equatorial distance $\langle R_{eq} \rangle$ corresponds to the average of the four equatorial distances while the axial distance $\langle R_{ax} \rangle$ is the average of the two axial ones. (c) Correlation diagram of the Cu$^{2+}$ $d$ levels in $O_h$ and $D_{4h}$ symmetries. Arrows indicate the three observed electronic transitions $E_1$, $E_2$ and $E_3$. The tetragonal splitting of the parent octahedral $e_g$ and $t_{2g}$ orbitals is $\Delta_e = E_1$ and $\Delta_t = E_3 - E_2$, respectively.

**Figure 2.** Variation of the optical absorption spectrum of CuWO$_4$ with pressure. Ticks show the placement of the three absorption bands.

**Figure 3.** (color online). Polarized absorption spectrum of CuWO$_4$ along the two extinction directions π and σ in the (100) plane. Each spectrum was obtained at ambient conditions in two polarizations allowing to resolve the $E_2$ band. The Gaussian fits to the three electric-dipole crystal-field transitions of Cu$^{2+}$ at ambient pressure ($D_{4h}$ symmetry) is included. Top figure shows the $E_1$ spectral subtraction to resolve $E_2$ and $E_3$.

**Figure 4.** (color online). Pressure dependence of the transition energy for $E_1$, $E_2$ and $E_3$ (on top). The lines represent least-square fits to the experimental data. The large error bars represent uncertainty due to masking effects. Hollow colored circles represent $E_2$ (blue) and $E_1$ (red) at the triclinic and monoclinic phases, respectively. Bottom-left part of the figure shows the variation of the JT distortion $Q_\theta$ with pressure [13]. The variation of the $e_g$- and $t_{2g}$- orbital splittings, $\Delta_e$ and $\Delta_t$ , as a function of the JT distortion, $Q_\theta$, in the low-pressure-triclinic and high-pressure-monoclinic phases are shown in the bottom-right figure.

Figure 1.

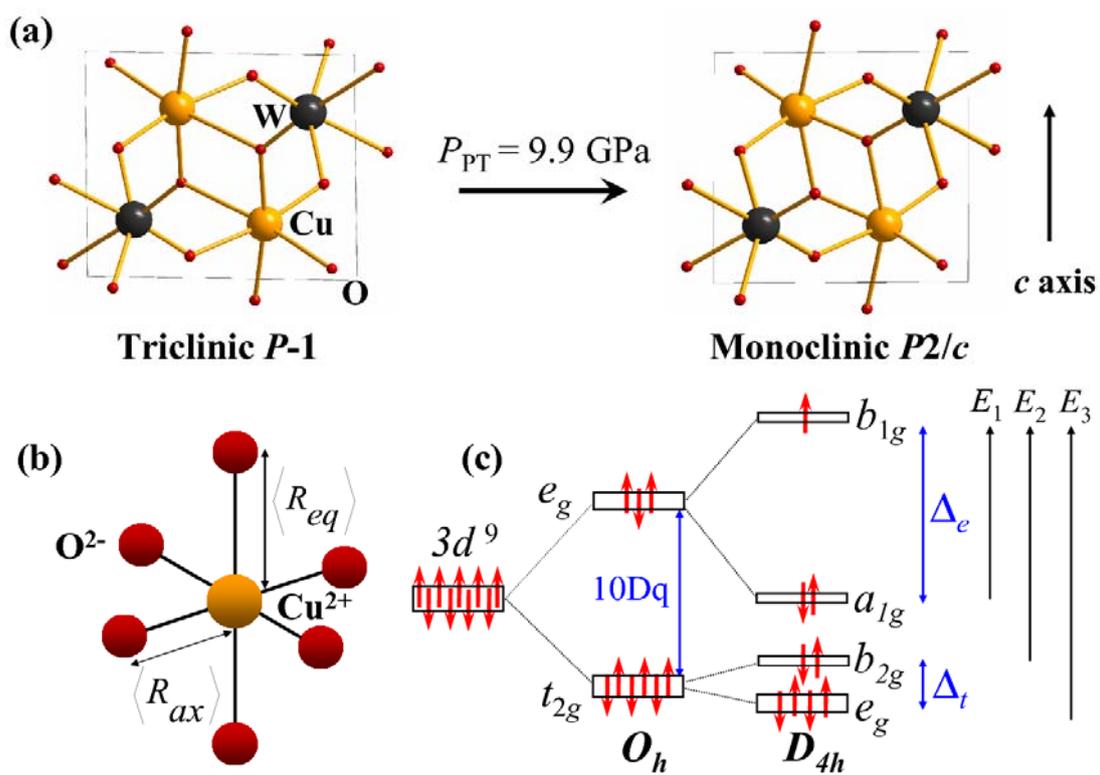

Figure 2.

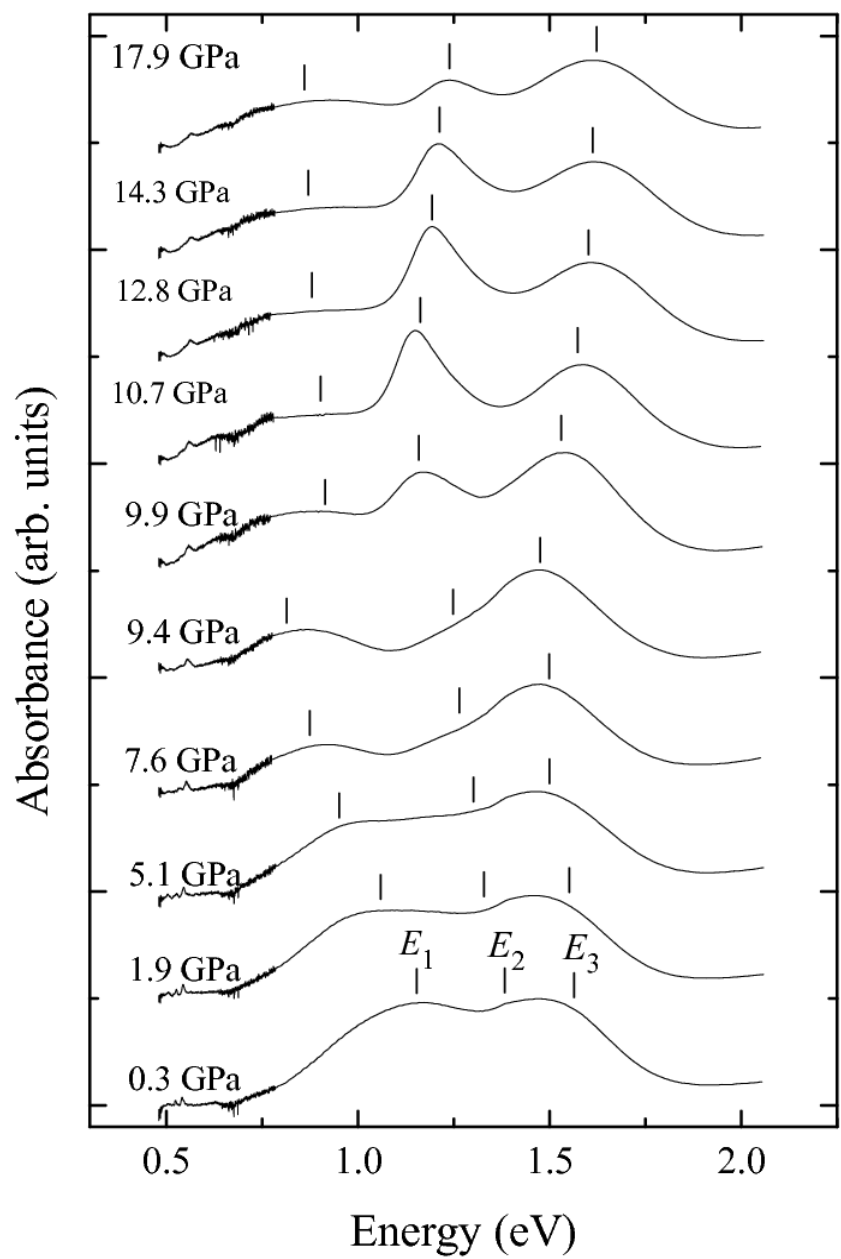

Figure 3.

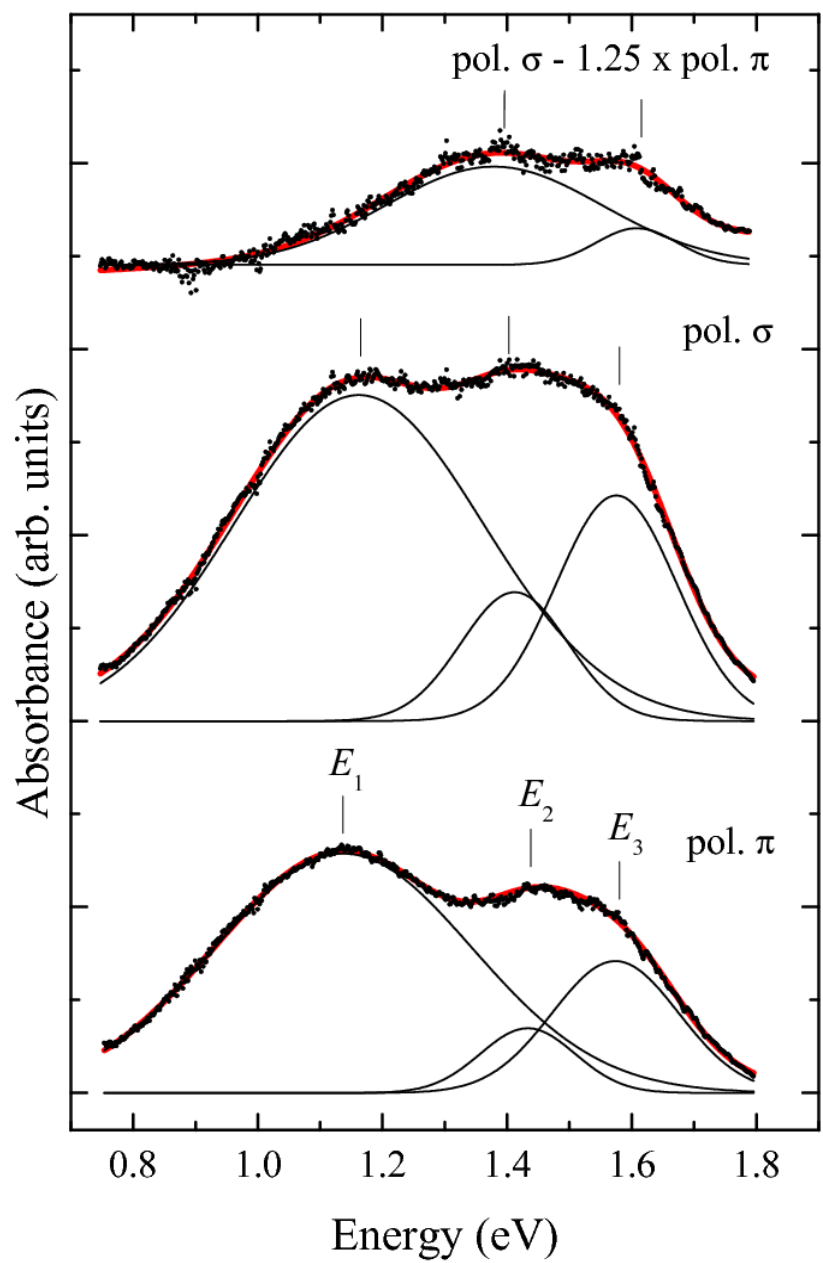

Figure 4.

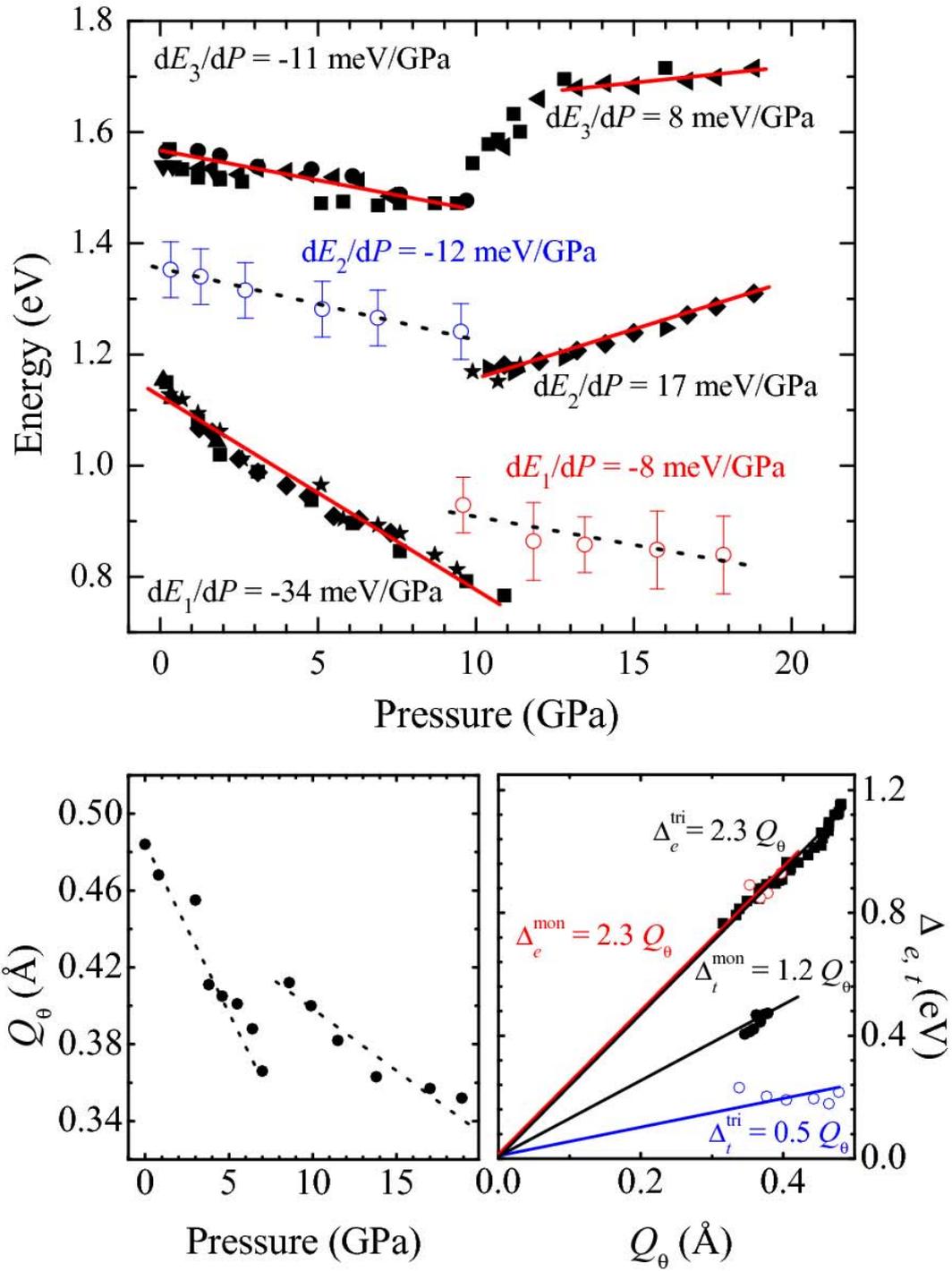